\documentclass[a4paper,10pt,twocolumn,twoside,notitlepage,final]{article}

\usepackage{amsmath}
\usepackage{bm}
\usepackage{graphicx}
\usepackage{fancyhdr}
\usepackage{Interface}
\makeatletter
\renewcommand{\@biblabel}[1]{#1}
\makeatother

\begin{document}

\title{Interlimb neural connection is not required for gait transition in quadruped locomotion}
\author{Atsushi Tero$^{\textbf{1}}$, Masakazu Akiyama$^{\textbf{2}}$, Dai Owaki$^{\textbf{3}}$, Takeshi Kano$^{\textbf{3}}$, Akio Ishiguro$^{\textbf{3,4}}$, and Ryo Kobayashi $^{\textbf{4,5}}$
\thanks{Author for correspondence (tero@imi.kyushu-u.ac.jp ).}}
\affiliation{
$^{1}$ Institute of Mathematics for Industry, Kyushu University, 744 Motooka, Nishi-ku, Fukuoka 819-0395, Japan.
$^{2}$ Research Institute for Electronics Science, Hokkaido University, N12W7, Kita-Ward, Sapporo, 060-0812, Japan.
$^{3}$ Research Institute of Electrical Communication, Tohoku University, 2-1-1, Katahira, Aoba-ku, Sendai 980-8577, Japan.
$^{4}$ Japan Science and Technology Agency CREST, 7 Goban-cho, Chiyoda-ku, Tokyo 102-0075, Japan.
$^{5}$ Faculty of Science, 1-3-1, Kagamiyama, Higashi-Hiroshima, 739-8526 Japan.
}
\abst{
Quadrupeds transition spontaneously to various gait patterns (e.g., walk, trot, pace, gallop) in response to the locomotion speed. The generation of these gait patterns has been the subject of debate for a long time. We propose a coupled oscillator model that is coupled with the physical interactions of the body. The results of this study showed that the gait pattern transitions spontaneously to walking/trotting/pacing/bounding in manner similar to that of real quadruped animals when the resonating portion of the body is changed according to the speed of leg movement. We also observed that pacing is expressed exclusively instead of trotting by changing the physical characteristics. In addition to leading to understanding of the principles of locomotion in living things, the coupled oscillator model proposed in this study is expected to lead to the creation of a legged robot that can select an energy-efficient gait and transition to it spontaneously.
}
\maketitle
\thispagestyle{empty}

%%%%%%%%%%%%%%%%%%%%%%%%%%%%%%%%%%%%%%%%%%%%%%%%%%%%%%%%%%%%%%%%%%%%%%%%%%%%%%%%
%%%                             1. INTRODUCTION                              %%%
%%%%%%%%%%%%%%%%%%%%%%%%%%%%%%%%%%%%%%%%%%%%%%%%%%%%%%%%%%%%%%%%%%%%%%%%%%%%%%%%
\section{INTRODUCTION}
\begin{figure*}[tb]
  \begin{center}
    \includegraphics[keepaspectratio=true,height=80mm]{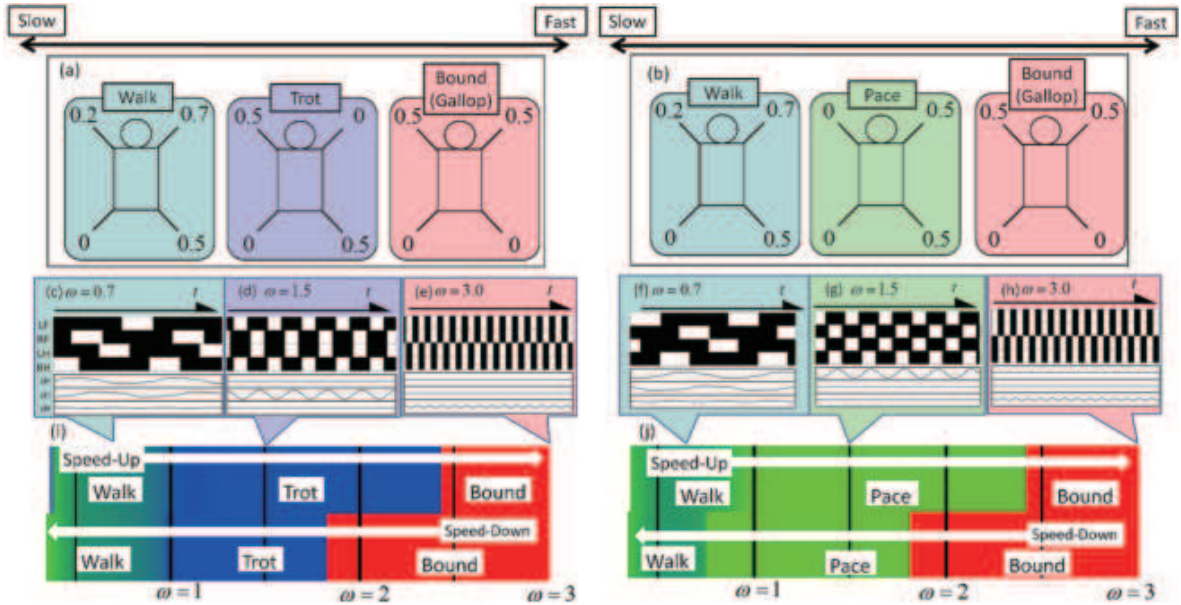}
  \end{center}
  \caption{Gaits of (a) horses and (b) camels. At low speeds, the gait is changed from a walk to trotting (pace) and bounding. (c)-(h) Results of numerical calculations for each parameter when the phase $\theta_i$ of a random leg was taken as the initial value and after sufficient time had elapsed $t=[9970, 10000]$. LF, RF, LH, RH each represent the respective leg $i$ and shows if it is a "stance" ($\sin\theta_i\le 0$, shown in black) or "swing" ($\sin\theta_i> 0$, shown in white ). (c)-(e) show the results of the numerical calculation when $\gamma_P=2.0$ and $\gamma_T=1.8$. (c), (d), and (e) show that a gait similar to that of a real horse (walk, trot and bound) can be reproduced when $\omega=0.7$, $\omega=1.5$, and $\omega=3.0$ respectively. (f)-(h) show the numerical calculations when $\gamma_P=1.8$ and $\gamma_T=2.0$. (g) shows that a gait similar to that of a camel can be reproduced where the trot of a horse is replaced by the gait. (i) and (j) show the simulation results when $\omega$ was slowly increased and decreased in the range of $[0.35,3.00]$. The upper section of (i) shows the numerical calculations when $\omega$ was increased by $\omega=0.0001t+0.35(0<t<26500)$, while the lower section of (i) shows the numerical calculations when $\omega$ was reduced by $\omega=5.65-0.0001t(26500<t<53000)$. The phase differences between the fore and hind legs, diagonally opposite legs, and left and right legs are marked in red, green, and blue, respectively (e.g., when the fore and hind legs are bounding simultaneously, they are marked in red). The gait transitions to walking, trotting, and bounding are the same as in real life. In all of the numerical calculations, $g=0.5,\gamma_B=4.0,k_P=3.9,k_T=4.0$ and $k_B=16.0$. Real quadruped animals such as monkeys move their legs in the following sequence: left fore (LF) $\to$ left hind (RH) $\to$ right fore (RF) $\to$ right hind (RH). Horses move their legs in the following sequence: left hind (LH) $\to$ left fore (LF) $\to$ right hind (RH) $\to$ right fore (RF). There is symmetry between the fore legs and hind legs. In this simulation, any parameter may be selected depending on whether an initial value or noise is given. By introducing asymmetry in the initial value, the selection can be fixed, and the sequence can be determined.}
  \label{fig:FigSimResultsAll.eps}
\end{figure*}

Quadrupeds transition spontaneously to various gait patterns (e.g., walking, trotting, pacing, galloping) in response to the locomotion speed \cite{muybridge1888} \cite{hildebrand1965}. Animals like horses usually transition sequentially from a walk to a trot and then gallop (Fig. \ref{fig:FigSimResultsAll.eps}a), whereas animals like camels transition from a walk to a pace and then gallop (Fig. \ref{fig:FigSimResultsAll.eps}b). Based on the spontaneous gait transitions of decerebrated cats (i.e., the neural connection between the spinal cord and brain is surgically severed) given in\cite{shilk1966}, the intraspinal neural network called the central pattern generator (CPG) and musculoskeletal properties of the limbs are thought to play an important role in gait transitions rather than the brain. Although various CPG network models have been proposed\cite{righetti2008}\cite{golbitsky1999}, they do not provide a clear explanation for the mechanism of formation of many gait transitions.
In contrast, theories that focused on musculoskeletal properties have reported that quadrupeds achieve optimal energy efficiency by gait transitions\cite{hoyt1981} and that oscillation of the legs and the physical interaction of the body are very important to achieving energy efficiency during walking and running \cite{Ahlborn2002}. In horse racing, the body (trunk) of the galloping horse oscillates back and forth in accordance with the phase of the legs. The race record becomes better when the horse rider oscillates in the reverse phase of the horse's phase \cite{Pfau2009}. Riders are also known to feel "motion sickness" due to the left and right movements of the camel's body caused by its walk and pace gaits. In this manner, the movement of the legs and oscillation of the body are closely related. Experimental modal analysis carried out on quadruped models has indicated the existence of vibration modes that generate pacing, trotting, and bounding gaits for the bodies of quadruped animal \cite{Kurita2008}. Other spontaneous gait transitions of walking and trotting have also been confirmed during passive walking (i.e., sensory and motion forces are absent) \cite{Osuka1985}. Although the results of many experiments have suggested the importance of physical interaction between the body parts during gait transition, this has not be proved completely, and only a limited reproduction of gait transition has been achieved so far.
In this article, we explain the mechanism of gait transition by using a simple model that accounts for the physical characteristics of the body. Specifically, we constructed a mathematically modeled oscillator that describes the behavior of each leg and caused interactions in the body to reproduce gait transitions similar to those of quadruped animals. Based on our results, we propose that the gait transition of quadrupeds can be generated simply by the physical interaction of body parts without any neural coupling between the legs.
% }'Ì'}"ü
\section{MATHEMATICAL MODEL}
\begin{figure*}[h]
  \begin{center}
    \includegraphics[keepaspectratio=true,height=100mm]{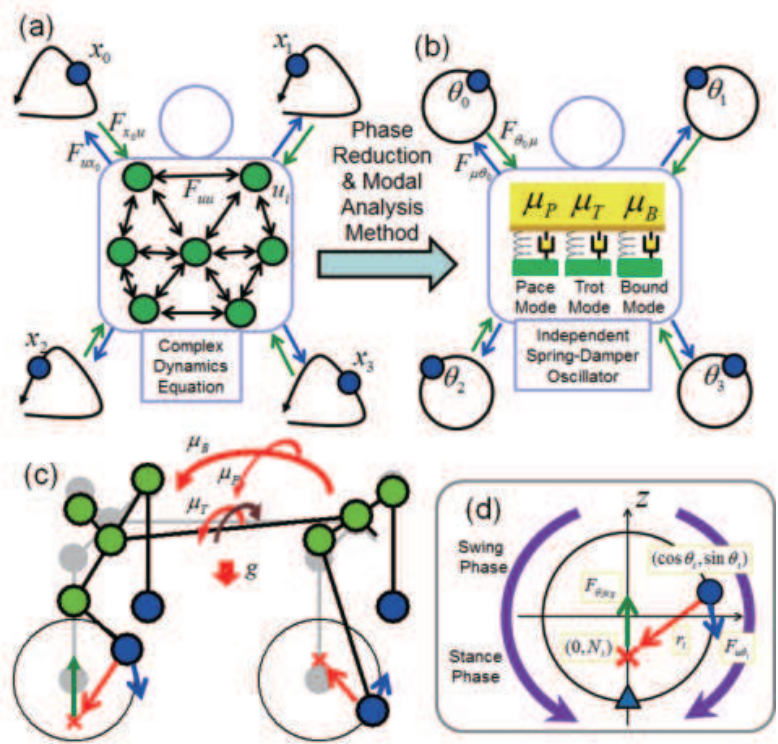}
  \end{center}
  \caption{ (a) Top view of the quadruped and outline of the mathematical model. In this model, each leg of the quadruped is represented by a phased oscillator to examine its interaction with the body. (b) Analysis, particularly of the complex dynamics of the body. We carried out a modal analysis and focused only on the three sets of complex eigenvalues measured against the eigenplanes. (c) Concrete implementation image of this model. This model can be realized by mounting four independent forces (legs) onto the four corners of the H-shaped body comprising the shoulder, spine, and hip. (d) Enlarged view of the left fore leg. In the above diagram, $\mu_P$ represents a rotation in the roll direction, $\mu_T$ represents the twist of the spine/hip, and $\mu_B$ represents the pitch direction towards the body. $u_0$ and $u_2$ represent the shifts of the left fore leg and left hind leg, respectively, in the $z$ direction from the standard value. 
$u_0=-g-\mu_P-\mu_T-\mu_B$,$u_2=-g-\mu_P+\mu_T+\mu_B$
Assume that, as in the above diagram, $g=0.1,\mu_P=0.1,\mu_T=0.3,\mu_B=0.2$. Then, $u_0=-0.7$ and $u_2=0.3$, and the standard position of each leg moves accordingly to become $N_0=-0.7$ and $N_2=0.3$ (indicated by the red "X"). The red arrows represent the shift $\mbox{\boldmath $r$}_i=(-\cos\theta_i,N_{i}-\sin\theta_i)$ between the leg and body. During this time, phase $\theta_0$ of the left fore leg goes into the "stance" phase, and phase $\theta_2$ of the left hind leg goes into the "swing" phase (marked by blue arrows). The purple arrows show the influence of $F_{u \theta_i}$ on $\theta_i$ when $N_i<0$. When $N_i<0$ and $0<\theta_i<\frac{\pi}{2}$ or $\frac{3\pi}{2}<\theta_i<2\pi$, $F_{u \theta_i}$ becomes less than $0$, i.e., $F_{u \theta_i}<0$. When $\frac{\pi}{2}<\theta_i<\frac{3\pi}{2}$, $F_{u \theta_i}$ becomes greater than $0$, i.e., $F_{u \theta_i}>0$. Thus, it is easier for that leg (indicated by the blue triangle) to become the stance phase. Conversely, when $N_i>0$, that leg can easily go into the swing phase. As a reaction to this, $\mu_P,\mu_T,$ and $\mu_B$ are influenced so that the shifts $u_0, u_2$ are small (marked by the green arrow).
}
  \label{fig:FigModel.eps}
\end{figure*}

We describe the mathematical model for a quadruped here. We use non-dimensional formulae throughout the explanation. We constructed mathematical models for the legs and body (trunk), and we discuss the interaction between them here (Fig. \ref{fig:FigModel.eps}a). For convenience, each leg is numbered as follows: left foreleg $\to$ 0, right foreleg $\to$ 1, left hind leg $\to$ 2, right hind leg $\to$ 3. The state variable of the $i$th leg is expressed as $\Vec x_{i}=^t (x_{i0},x_{i1},x_{i2},\cdots )$. The physical state variable of the body is expressed as $\Vec u=^t (u_{0},u_{1},u_{2},\cdots )$. The relationship between these variables is given below.
\begin{eqnarray}
\dot{\Vec x}_{i}=F_{xx_{i}}+F_{u x_{i}}\label{Eq:Leg1}\\
\dot{\Vec u}=F_{uu}+\sum_{i=0}^3 F_{x_i u}\label{Eq:Body1}
\end{eqnarray}
$F_{xx_{i}}(\Vec x_{i})$ represents the dynamics of the $i$th leg. $F_{u x_{i}}$ represents the external force on the $i$th leg from the body. $F_{uu}$ represents the dynamics of the body, and $F_{x_i u}$ represents the external force from the $i$th leg to the body. We assumed that direct interaction between the legs is absent (no neural interaction) and that the gait transition is influenced only by physical interaction.\\
\\
We simplified these equations by applying a phase reduction to state variable $\Vec x_{i}$ of the leg and modal analysis for the body state variable $\Vec u$ (Fig. \ref{fig:FigModel.eps}b). Each leg of quadruped animals performs the following movements in a cyclical manner while walking and running:Swing$\to$Contact$\to$Stance$\to$Kick$\to$Swing. \cite{grillner1985} shows that autonomous neural oscillations exist within lamprey eels. Quadruped animals should also incorporate similar leg movements due to autonomous neural oscillations \cite{righetti2008}\cite{golbitsky1999}. On the other hand, the legs of stick insects carry out cyclical movements based on their physical condition and neural reflexes \cite{Dean1999}. We are not arguing over the mechanism with which quadruped animals carry out cyclical actions; however, in either case we can describe the neural and physical states during phase $\theta_i$ of one cycle. When $\sin\theta_i>0$, then leg $i$ is in a "swing" phase; when $\sin\theta_i\le 0$, then leg $i$ is in a "stance" phase. We assumed that the height of each leg can be approximated to $\sin\theta_i$. Based on the above, phase reduction can be performed on the state variables for each leg using (\ref{Eq:Leg1},\ref{Eq:Body1}); these equations can be rewritten as given below.\\
\begin{eqnarray}
\dot{\theta_i}=\omega +F_{u \theta_i} +\xi\label{Eq:Leg2}\\
\dot{\Vec u}=F_{uu}+\sum_{i=0}^3 F_{\theta_i u}\label{Eq:Body2}
\end{eqnarray}
$\omega$ is the angular velocity of the leg; when a quadruped tries to run fast, it takes a larger $\omega$ value. $F_{u \theta_i}(\Vec u)$ represents the influence the body has on each leg; $F_{u x_{i}}$ represents the transformed variable, while $\xi(t,i)$ represents noise. When the influence of the body on the leg is not taken into consideration ($F_{u \theta_i}=0$), the leg moves at a constant angular velocity, and the support ratio (duty factor) becomes $0.5$.\\
\\
We reduced the dimensions of the state variable for body $\Vec u$ by carrying out modal analysis (Fig. \ref{fig:FigModel.eps}b). Modal analysis can be explained in simple terms as an engineering technique used to understand the properties of complex structures by means of linear approximation, variable transformations, and analysis in each eigenspace in order to identify and ignore the variables with high damp. In quadruped animals, the damp is known to be faster than the skeletal forms for all eigenplanes other than those that take gait patterns like pacing, trotting, and bounding \cite{Kurita2008}. Thus, we focused only on the bases of these eigenplanes $\Vec\mu=^t(\mu_P,\dot{\mu_P},\mu_T,\dot{\mu_T},\mu_B,\dot{\mu_B})$. Even though various factors can be considered when discussing the movement of a body that paces, trots, and bounds, we discuss the body's rotation in the roll direction $\mu_P$, twist of the spine $\mu_T$, and rotation in the direction of the pitch towards the body $\mu_B$ as examples (Fig. \ref{fig:FigModel.eps}c).
\begin{eqnarray}
\dot{\theta_i}=\omega +F_{\mu \theta_i} +\xi\label{Eq:Leg3}\\
\ddot{\mu_X}=-\gamma_X \dot{\mu_X}-\kappa_X \mu_X+\sum_{i=0}^3 F_{\theta_i \mu_X}\label{Eq:Body3}
\end{eqnarray}
Here, $X \to P,T,B$ and $F_{\mu_X \theta_i}(\theta_i,\Vec\mu)$,$F_{\theta_i \mu_X}(\theta_i,\Vec\mu)$ are the transformations of $F_{u_X\theta_i }$ and $F_{\theta_i u_X}$ (Fig. 2b).\\
We now discuss the external force $F_{\mu \theta_i}=\Vec Z(\theta_i)\cdot \Vec p_i(\Vec \mu)$ on the leg from the body. $\Vec Z(\theta_i)$ is the phase sensitivity function, and we utilize $\Vec Z(\theta_i)=(-\sin\theta_i,\cos\theta_i)$. Assuming that the external force falls only at vertical angles, assume $\Vec p_i(\Vec\mu)=(0,N_i(\Vec \mu))$. We then substitute all of this into (\ref{Eq:Leg3}) to give us the following formula:
\begin{eqnarray}\label{Eq_theta1}
\dot{\theta_i}=\omega +N_{i}\cos\theta_i+\xi\label{Eq:Leg4}
\end{eqnarray}
We assume that $N_i$ can be expressed as a linear expression of $\mu_P,\mu_T$ and $\mu_B$:
$$N_i(\Vec\mu)=-g+n_{iP}\mu_P+n_{iT}\mu_T+n_{iB}\mu_B$$

where $n_{iP},n_{iT},n_{iB},g$ are constants. The external force from $\mu_P$ causes the leg to be in the "pace" gait. Thus, if we assume that $n_{0P}<0$, then $n_{2P},n_{1P},n_{3P}$ should be as given below.

$$n_{2P}<0,n_{1P}>0,n_{3P}>0$$

Therefore, let $n_{0P}=n_{2P}=-1$ and $,n_{1P}=n_{3P}=1$. Similarly, let $n_{0T}=n_{3T}=-1$ and $n_{1T}=n_{2T}=1$ and let $n_{0B}=n_{1B}=-1,n_{2B}=n_{3B}=1$. Then, $N_{i}$ becomes as given in the equations below.
\begin{eqnarray}
N_{0}=-g-\mu_P-\mu_T-\mu_B \label{Eq_theta2} \\
N_{1}=-g+\mu_P+\mu_T-\mu_B \label{Eq_theta3} \\
N_{2}=-g-\mu_P+\mu_T+\mu_B \label{Eq_theta4} \\
N_{3}=-g+\mu_P-\mu_T+\mu_B \label{Eq_theta5}
\end{eqnarray}
By substituting (\ref{Eq_theta2}-\ref{Eq_theta5}) into (\ref{Eq:Leg3}), the following equations on the progress of the phase of the leg can be derived.
\begin{eqnarray}\label{Eq_theta}
\dot{\theta_0}=\omega +(-g-\mu_P-\mu_T-\mu_B)\cos \theta_0 +\xi\\
\dot{\theta_1}=\omega +(-g+\mu_P+\mu_T-\mu_B)\cos \theta_1 +\xi\\
\dot{\theta_2}=\omega +(-g-\mu_P+\mu_T+\mu_B)\cos \theta_2 +\xi\\
\dot{\theta_3}=\omega +(-g+\mu_P-\mu_T+\mu_B)\cos \theta_3 +\xi
\end{eqnarray}
In the example given in Fig. \ref{fig:FigModel.eps}c, the up and down shift from the standard value of the left shoulder is $-g-\mu_P-\mu_T-\mu_B$. If $g,\mu_P,\mu_T$ and $\mu_B$ become large, this deviation alone goes down. The results of various experiments have shown that horses have a gait transition that aids in reducing the localized load on that tendon and muscle \cite{Farley2010}. We consider the shift $\Vec r_i=(-\cos\theta_i,N_{i}-\sin\theta_i)$ between the standard value $(0,N_{i})$ for the joint between the body and leg and the actual state of the leg $(\cos\theta_i,\sin\theta_i)$ to be proportional to the load, and we propose that the leg is influenced so that this shift becomes small. In other words, we consider $\Vec r_i$ to be a virtual spring of natural length $0$ and spring constant $1$, and we propose that $\theta_i$ varies with its righting moment \cite{umedachi2010}. Then, $F_{\mu \theta_i}$ can be represented as given below, and we can derive results similar to that of (\ref{Eq:Leg4}).
$$\displaystyle{F_{\mu \theta_i}=-\frac{1}{2}\frac{\delta  |\mbox{\boldmath $r$}_i|^2}{\delta \theta_i}=N_{i}\cos\theta_i }$$

If $N_{i}<0$, that leg can easily move to a "stance" phase easily; if $N_i>0$, that leg can easily move to a "swing" phase (purple arrow in Fig. \ref{fig:FigModel.eps}d). As a result, for a stationary quadruped animal that is standing upright ($\omega=\mu_P=\mu_T=\mu_B=0$), $\dot{\theta_i}$ becomes $\dot{\theta_i}=-g \cos \theta_i +\xi$, and all of the legs converge to the "stance" phase ($\theta_i\to \frac{3\pi}{2}$). If we assume that $\mu_P\gg \omega,g,|\mu_T|$ and $|\mu_B|$, then $\theta_0$ and $\theta_2$ become $\frac{3\pi}{2}$ ($\theta_0,\theta_2\to \frac{3\pi}{2}$), and $\theta_1$ and $\theta_3$ become $\frac{\pi}{2}$ ($\theta_1,\theta_3\to \frac{\pi}{2}$): according to the definition of $\mu_P$, the gait becomes a "pace."

Next, we consider the influence of the leg $F_{\theta_i \mu_X}(\theta_i,\mu)$ on the body. Similar to the discussion regarding the influence on the leg by the body ($F_{u \theta_i}$), for $F_{\theta_i \mu_X}(\theta_i,\mu)$ we also propose that the leg is influenced so that the shift between the body and leg $\Vec r_i$ is small (green arrow in Fig. \ref{fig:FigModel.eps}d).
\begin{eqnarray}\label{Eq:Body2}
F_{\theta_i \mu_X}=-\frac{1}{2}\frac{\delta |\Vec r_i|^2}{\delta \mu_X}
\end{eqnarray}
Based on (\ref{Eq:Body1}) and (\ref{Eq:Body2}), the following equations can be derived. 
\begin{eqnarray}
\begin{split}
\ddot{\mu_P}=&-\gamma_P \dot{\mu_P}-k_P \mu_P\\
 & -\sin\theta_0+\sin\theta_1-\sin\theta_2+\sin\theta_3\label{Eq:Body3_P}\\
\end{split}
\end{eqnarray}
\begin{eqnarray}
\begin{split}
\ddot{\mu_T}=&-\gamma_T \dot{\mu_T}-k_T \mu_T\\
&-\sin\theta_0+\sin\theta_1+\sin\theta_2-\sin\theta_3\label{Eq:Body3_T}\\
\end{split}
\end{eqnarray}
\begin{eqnarray}
\begin{split}
\ddot{\mu_B}=&-\gamma_B \dot{\mu_B}-k_B \mu_B\\
&-\sin\theta_0-\sin\theta_1+\sin\theta_2+\sin\theta_3\label{Eq:Body3_B}
\end{split}
\end{eqnarray}
Here, $k_X$ is $k_X=\kappa_X-4(X\to P,T,B)$).
\section{SIMULATION RESULTS}
Here, we discuss the simulation results for the model described in the previous section. For example, a real horse transitions to various gaits like walking, trotting, and bounding depending on the speed. Here, the speed of the quadruped animal is expressed by the characteristic angular velocity $\omega$ of each leg. In the numerical experiments, when $\omega$ was small, the walking gait was reproduced (Fig. \ref{fig:FigSimResultsAll.eps}c). When the locomotion speed is faster than a walk, a real horse would transition to a trot. Similarly, when the value of $\omega$ was increased in the simulation, the gait became a trot (Fig. \ref{fig:FigSimResultsAll.eps}d). During this time, the twist of the body $\mu_T$ was in tune with the movement of the legs, and its oscillation became large. This gait manifested because of the resonance between $\mu_T$ and the oscillation of the leg. When the locomotion speed is faster than a trot, a real horse then transitions to a gallop or bound. In the simulation, increasing the value of $\omega$ also caused the gait to change from a trot to a bound (Fig. \ref{fig:FigSimResultsAll.eps}e). Similar to the trot, this gait manifested because of the resonance between $\mu_B$ and the oscillation of the leg. When $\omega$ was very low, walking did not happen (The legs stop.); when $\omega$ became too large, the oscillation of the leg became very fast with respect to the body movement. Thus, a stable gait pattern could not be generated. On the other hand, a pace did not manifest with the parameters given in Figs. \ref{fig:FigSimResultsAll.eps}c-e, similar to a real horse. However, a camel has different physical characteristics from a horse and paces instead of trotting. By changing the physical parameters, the gait became a pace instead of a trot during the simulation (Fig. \ref{fig:FigSimResultsAll.eps}g). As a result, gait transitions (walking, pacing, and bounding) similar to those of a camel were reproduced (Figs. \ref{fig:FigSimResultsAll.eps}f-h).\\
\\
Next, we took the initial conditions to be $\theta_i=\frac{3\pi}{2} ( i = 0,1,2,3)$ and applied an angular velocity of $\omega=0.35$. We then slowly increased $\omega$ over time (upper section of Fig. 1i). After $\omega$ reached $\omega=3.0$, we then slowly reduced $\omega$ (lower section of Fig. \ref{fig:FigSimResultsAll.eps}i). Thus, we were able to reproduce the walking, trotting, and bounding gaits similar to the gaits produced during the acceleration and deceleration of a real quadruped animal. Hysteresis was confirmed during these gait transitions. Similarly, the gait transitions were also reproduced for a camel (Fig. \ref{fig:FigSimResultsAll.eps}j).
\section{DISCUSSION}
The biological meaning of a gait transition can be clearly explained in terms of energy consumption rate (oxygen consumption rate) \cite{hoyt1981} and reduction of body weight or injury \cite{Farley2010}. Thus, if the quadruped determines the gait transition and characteristic angular velocity $\omega$ of the leg, then it can be thought of as a coupled oscillator that spontaneously selects the optimal phase difference suitable for energy efficiency and load. For example, static stability is necessary for walking at a low speed; in real quadruped animals, the duty factor is high while walking. In this model, the duty factors of each leg while walking, trotting, and bounding are $0.69, 0.61$, and $0.60$ respectively (Figs. \ref{fig:FigSimResultsAll.eps}c-e). Therefore, results can be achieved even if the duty factors are not given explicitly.\\
\\
When $g$ was taken as $0$ ($g=0$) and the body weight was ignored, walking was not a stable solution. When $g>0$, a solution where more than two of $\mu_P,\mu_T$ and $\mu_B$ oscillate became stable. However, the results of the experimental modal analysis \cite{Kurita2008} ($k_P\simeq k_T<k_B$) showed almost no change in $\mu_B$ with a high righting moment $k_B$. As a result, the solution for a walk is the same as that of a pace and trot where $\mu_P$ and $\mu_T$ coexist together (Figs. \ref{fig:FigSimResultsAll.eps}c and f). Thus, when $\omega$ is high, $k_P, k_T$ and $k_B$ influence the natural frequency and produce an exclusive pace, trot, and bound that correspond to the high-speed region. However, in the low-speed region where $\omega$ is low, $k_P, k_T$ and $k_B$ cause the manifestation of a walk gait that is a common solution for a pace and trot. Although the diagonality (phase difference between the fore and hind legs) was taken as $0.2$ in Figs. 1a and b, the diagonality varies with each species for real quadruped \cite{hildebrand1965}. This model can also reproduce various phase differences depending on the values of $k_P, k_T$, etc.\\
\\
In this study, only the musculoskeletal system of a quadruped was taken as the state variable $\Vec u$. However, real-life quadruped animals may carry out gait transition by using both their nervous system and physical body. In this case, both the nervous system and physical body may need to be taken as the state variable $\Vec u$. 
The order parameters of oscillators represented in models like the Kuramoto model have a behavior equivalent to the Stuart-Landau equations \cite{Ott2008}, and the order parameters show damping behavior during asynchronous operation of oscillators. Consequently, the order parameters of the neural oscillators (called CPG) observed in Lamprey eels may behave similar to (\ref{Eq:Body3_P}-\ref{Eq:Body3_B}) in response to the synchronous and asynchronous modes of the oscillators. Thus, it is necessary to carefully reconsider and check if the gait transition of quadruped animals is based on the physical body or neural system.

\section{SUMMARY}
We performed a modal analysis on the physical interaction of the body and constructed a coupled oscillator model. In this model, gait transitions (walking, trotting, and bounding) similar to those of real quadruped animals such as horses were reproduced simply by varying the angular velocity of the leg $\omega$. We also changed the gait transitions to be similar to those of other quadruped animals like camels (walk, pace and bound) by slightly changing the physical parameters. During this activity, the duty factor for each gait was also replicated. Based on these results, the gait transition and many of the associated phenomena in quadrupeds may be excited by physical interactions.

{\small
\section*{Acknowledgement}
This work was supported by JSPS KAKENHI Grant Number 22686040.
}

\end{document}